\documentclass[12pt]{article}

\usepackage{axodraw}
\usepackage{graphicx}
\usepackage{defsign,defref}

\newcommand \ket[1]{\left\vert\, {#1} \, \right>}

\newcommand{\bea}{\begin{eqnarray}}
\newcommand{\eea}{\end{eqnarray}}
\newcommand{\simgt}{\hbox{ \raise3pt\hbox to 0pt{$>$}\raise-3pt\hbox{$\sim$} }}
\newcommand{\simlt}{\hbox{ \raise3pt\hbox to 0pt{$<$}\raise-3pt\hbox{$\sim$} }}
\newcommand \vc[1]{{\bf {#1}}}

\begin{document}
\begin{titlepage}
\title{
~\vspace{1cm}\\
Probe of {\it CP} Violation in $e^+e^- \to t\bar{t}$\\
Near Threshold%
\thanks{Talk given
at the Kiken Meeting:
``New Perspectives in Elementary Particle Physics'',
Kyoto, Japan, July 17 - 20, 2000.}
\vspace{2cm}
}
\author{Y.~Sumino
\\ \\ Department of Physics, Tohoku University\\
Sendai, 980-8578 Japan
}
\date{}
\maketitle
\thispagestyle{empty}
\vspace{-4.5truein}
\begin{flushright}
{\bf TU--597}\\
{\bf July 2000}
\end{flushright}
\vspace{3.0truein}
\vspace{4cm}
\begin{abstract}
\noindent
{\small
We report our theoretical study on
how to probe the anomalous {\it CP}-violating couplings of
the top quark in the $t\bar{t}$ region at
future $e^+e^-$ linear colliders.
We focus on the unique role of the $t\bar{t}$ threshold region.
}
\end{abstract}
%\vfil

\end{titlepage}
  
\section{Introduction}

In this article we report our recent theoretical study \cite{jns} on 
how to probe {\it CP} violation in the top quark sector
at future $e^+e^-$ linear colliders
in the $t\bar{t}$ threshold region.

\begin{figure}[h]
  \hspace*{\fill}
    \includegraphics[width=5cm]{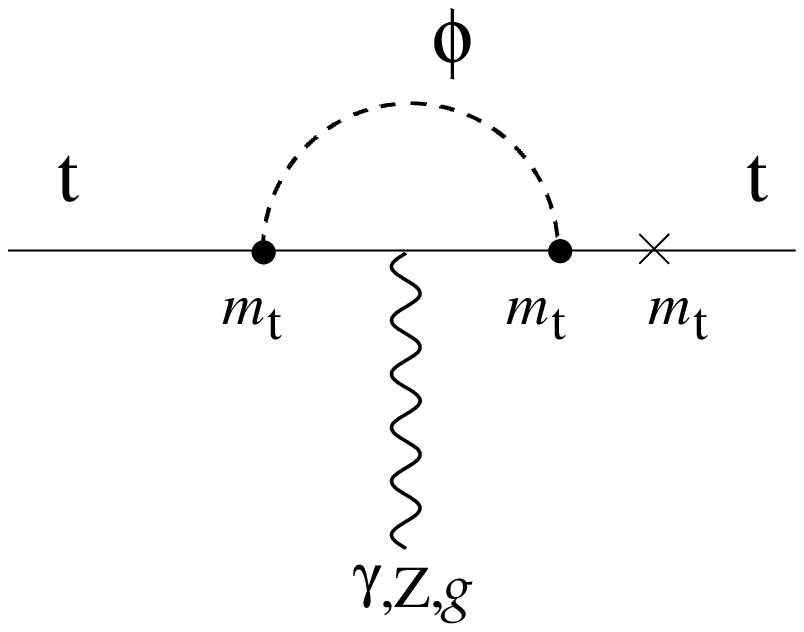}
  \hspace*{\fill}
%  \\
%  \hspace*{\fill}
\caption{\footnotesize
In two-Higgs-doublet models,
a neutral Higgs boson $\phi$ can violate {\it CP}
through the Yukawa interaction.
The top quark EDM is induced by an exchange of $\phi$
at one loop and is proportional to $m_t^3$.
One power of $m_t$ is necessary to flip chirality;
extra two powers come from the Yukawa interaction.
Since this is a one-loop effect, it is much larger than the
SM EDM; since it is proportional to $m_t^3$, it is 
strongly enhanced in comparison to the other quarks' EDMs.
      \label{THDM}
}
  \hspace*{\fill}
\end{figure}

Recently studies of various properties of the top quark have been started at 
Tevatron.
The
detailed properties will be investigated further
in future experiments at LHC and at future $e^+e^-$ linear
colliders.
Among various interactions of the top quark, 
testing the {\it CP}-violating interactions
is particularly interesting due to 
following reasons:
\begin{itemize}
\item
Within the Standard Model (SM),
{\it CP}-violation in the top quark sector is extremely small.
[The electric-dipole-moment (EDM) of a quark 
is induced first at three-loop level \cite{EDM-SM}.]
If any {\it CP}-violating effect is detected in the top quark sector 
in a near-future experiment, it immediately signals new physics.
\item
There can be many sources of {\it CP}-violation
in models that extend the SM, such as supersymmetric models, 
Leptoquark models, multi-Higgs-doublet models,
Extra-dimensions, etc.
Besides, the observed baryon asymmetry in the Universe suggests 
existence of {\it CP} violating mechanisms beyond the SM.
\item
In relatively wide class of models
beyond the SM, {\it CP} violation 
emerges especially sizably in the top quark sector.
A typical example is shown in Fig.~\ref{THDM}.
\end{itemize}

Let us state the set-ups of our analysis.
We consider 
{\it CP}-violating interactions of top quark with
$\gamma$, $Z$, and $g$.
In particular, we consider
the lowest dimension {\it CP}-odd effective operators:
\bea
{\cal L}_{\mbox{\scriptsize {\it CP}-odd}} =
- \frac{e d_{t\gamma}}{2m_t} 
( \bar{t} i \sigma^{\mu\nu} \gamma_5 t )
\partial_\mu A_\nu
- \frac{g_Z d_{tZ}}{2m_t} 
( \bar{t} i \sigma^{\mu\nu} \gamma_5 t )
\partial_\mu Z_\nu
- \frac{g_s d_{tg}}{2m_t} 
( \bar{t} i \sigma^{\mu\nu} \gamma_5 T^a t )
\partial_\mu G_\nu^a ,
\nonumber\\
 \sigma^{\mu\nu} \equiv {\textstyle \frac{i}{2}}
[ \gamma^\mu , \gamma^\nu ] ,
\label{effop}
\eea
where 
$e = g_W \sin \theta_W$ and
$g_Z = {g_W}/{\cos \theta_W}$.
These represent the interactions of $\gamma$, $Z$, $g$ with
the EDM, $Z$-EDM, 
chromo-EDM of top quark, 
respectively.\footnote{
The magnitudes of these EDMs are given by ${e d_{t\gamma}}/{m_t}$,~
${g_Z d_{tZ}}/{m_t}$,~ ${g_s d_{tg}}/{m_t}$,
respectively.
$d_{t\gamma}=1$ corresponds to $e/m_t \sim 10^{-16}~e \, {\rm cm}$,
etc.
}
Each of the interactions has $C = +1$ and $P = -1$.
As stated, the SM contributions to these couplings are extremely small,
$d_{t\gamma}^{\rm (SM)}, d_{tZ}^{\rm (SM)}, d_{tg}^{\rm (SM)} \sim 10^{-14}$.
Since we will not be able to detect them in near-future collider experiments,
we neglect the SM contributions below.
Our concern is in the anomalous couplings which are induced from
some new physics.
We assume that generally the couplings
$d_{t\gamma}$, $d_{tZ}$, $d_{tg}$
are complex where their imaginary parts may
be induced from some absorptive processes.

Many of the readers would be interested in the sensitivities to
these couplings expected in future experiments.
In Table~\ref{tab1} we summarize the results of 
the sensitivity studies performed
so far, including the results of our present study.
\begin{table}[t]
\begin{center}
\small
\begin{tabular}{l|l||c|c|c}
\hline
\multicolumn{2}{l||}{}
& $\delta d_{tg}$ & $\delta d_{t\gamma}$ & $\delta d_{tZ}$
\\ \hline 
\multicolumn{2}{l||}{LHC ~~~ (10~fb$^{-1}$)}
& $10^{-2}$ -- a few$\times 10^{-3}$ & - & -
\\
\hline
$e^+e^-$ LC & open top &
${\cal O}(1)$ & $10^{-1}$ -- a few$\times 10^{-2}$ &
$10^{-1}$ -- a few$\times 10^{-2}$
\\
\cline{2-5}
(50 fb$^{-1}$) & $t\bar{t}$ threshold &
$10^{-1}$ & $10^{-1}$ & $10^{-1}$ 
\\
\hline
\end{tabular}
\caption{\small
The results of studies of sensitivities to the anomalous couplings
expected in future
experiments.
For $e^+e^-$ linear colliders (LC),
``open top'' denotes the results of studies performed at $\sqrt{s}=500$~GeV.
These results are taken from \cite{SP92}-\cite{CEDM-epem-col}.
}
\label{tab1}
\end{center}
\end{table}
We may compare the sensitivities of 
experiments in the $t\bar{t}$
threshold region at $e^+e^-$
colliders with others.
The sensitivities to $d_{t\gamma}$ and $d_{tZ}$ are comparable
to those attainable in the open-top region at $e^+e^-$
colliders.
The sensitivity to $d_{tg}$ is worse than that expected at 
a hadron collider but
exceeds the sensitivity in the open-top region
at $e^+e^-$ colliders.

From this comparison one may find that our study at
$t\bar{t}$ threshold has little
impact on the study of {\it CP} violation and is not very interesting.
The present author, however, 
has a slightly different physics interest personally.
Although admittedly
it is better to have higher sensitivities to the anomalous
couplings $d_{tg}$, $d_{t\gamma}$, $d_{tZ}$, 
at the moment we do not know the sizes of these
couplings.
Therefore, I am interested more in the following questions
than merely in achievable sensitivities:
When any of the couplings happens to be sizable enough to be
detected in some experiment,
through what intriguing phonomena
can we detect the anomalous couplings?
And how can we
extract as much information on the couplings as possible?
In these respects, the $t\bar{t}$
threshold region has fairly rich physics contents.
I would like to describe our investigations from this viewpoint below.
So, please imagine a situation where any of the couplings happens
to be of order 10\% or larger and see what we can learn in that case.

\section{$t\bar{t}$ Threshold}

\subsection{Unique aspects}

When studying {\it CP} violation of the top quark,
unique aspects of the $t\bar{t}$
threshold region are:
\begin{itemize}
\item
The QCD interaction is enhanced in this region, hence
the cross section is sensitive to the top-gluon ($tg$) couplings.
We can study anomalous $tg$ couplings in a clean environment
in comparison to hadron colliders.
\item
In certain models (e.g.\ those in which a neutral Higgs boson is
exchanged between $t$ and $\bar{t}$ \cite{EDM-2HDM}),
induced top quark EDM and $Z$-EDM are enhanced near
the $t\bar{t}$ threshold.
\item
Since top quarks are produced almost at rest, 
one can reconstruct
the spin information of top quarks from 
distributions of their decay
products without solving detailed kinematics.
\end{itemize}

\subsection{Time evolution of the $t\bar{t}$ system}

\begin{figure}[tbp]
  \hspace*{\fill}
    \includegraphics[width=7cm]{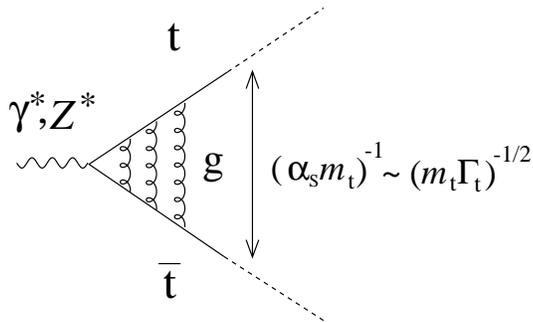}
  \hspace*{\fill}
  \\
  \hspace*{\fill}
\caption{\footnotesize
The time evolution of the $t\bar{t}$ system within the
SM.
      \label{SM}
}
  \hspace*{\fill}
\end{figure}

Let us first review the time evolution of 
$t$ and $\bar{t}$, pair-created in $e^+e^-$ collision 
just below threshold, within the SM (Fig.~\ref{SM}).
They are created close to each other
at a relative distance $r \sim 1/m_t$ 
and then spread apart non-relativistically.
When their relative distance becomes of the order of
the Bohr radius,
$r \sim ( \alpha_s m_t )^{-1}$,
they start to form a Coulombic boundstate.
When the relative distance becomes 
$r \sim (m_t \Gamma_t )^{-1/2}$,
where $\Gamma_t$ is the decay width of top quark,
either $t$ or $\bar{t}$ decays via electroweak interaction,
and accordingly the boundstate decays.
Numerically these two scales have similar magnitudes,
$( \alpha_s m_t )^{-1} \sim (m_t \Gamma_t )^{-1/2}$,
and are much smaller than the hadronization scale
$\sim \Lambda_{\rm QCD}^{-1}$.
Since gluons which have wavelengths much longer than the
size of the $t\bar{t}$ system cannot couple to 
this color singlet system, the strong interaction
participating in 
the formation of the boundstate
is dictated by the perturbative domain of QCD.
Due to this reason, we are able to compute the amplitude
from the first principles with 
order 5\% accuracy or better, even though
the QCD boundstates are involved.\footnote{
At the moment
only one exception is the normalization of the total
$t\bar{t}$ cross section, where we still have 10--15\%
theoretical uncertainty.
}
The spin and $PC$ of the dominantly produced boundstate  
are $J^{PC} = 1^{--}$.
Inside this boundstate: $t$ and $\bar{t}$ are in the 
$S$-wave state ($L = 0$);
the spins of $t$ and $\bar{t}$ are aligned to each other and
pointing to $e^-$ beam direction $\ket{\uparrow\uparrow}$
or to $e^+$ beam direction $\ket{\downarrow\downarrow}$ or
they are in a linear combination of the two states ($S=1$).

\begin{figure}[tbp]
  \hspace*{\fill}
  \begin{minipage}{5.0cm}\centering
    \includegraphics[width=4.5cm]{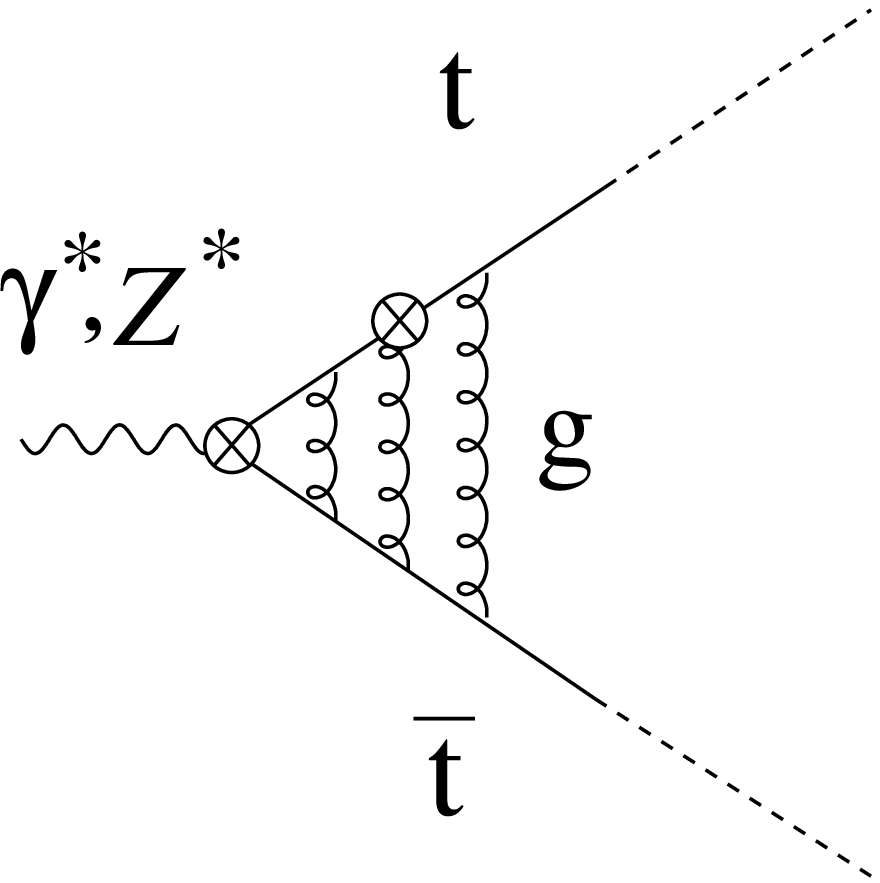}
\\
    (a)
  \end{minipage}
  \hspace*{\fill}
  \begin{minipage}{5.0cm}\centering
    \includegraphics[width=4cm]{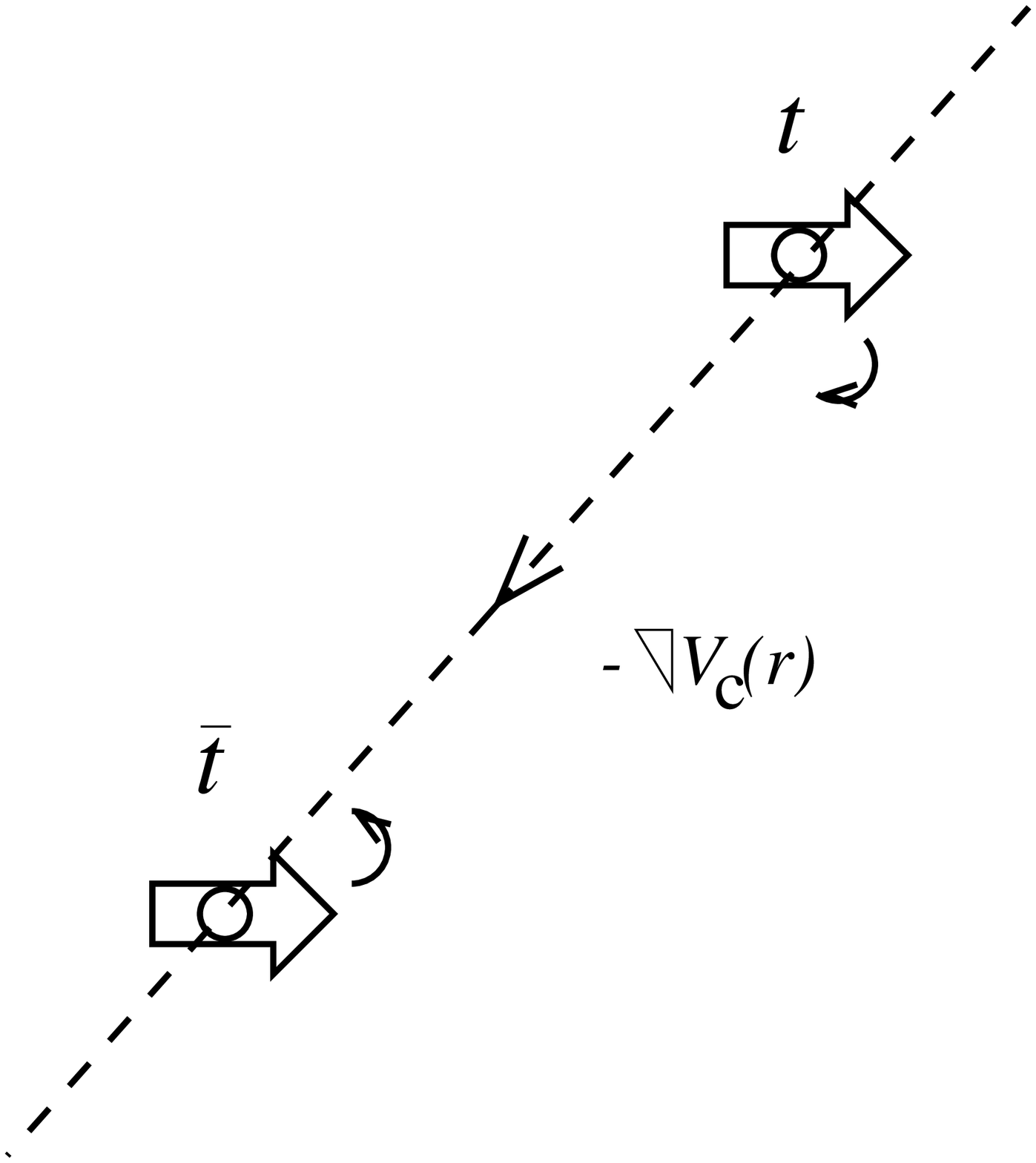}
\\
    (b)
  \end{minipage}
  \hspace*{\fill}
  \\
  \hspace*{\fill}
\caption{\footnotesize
(a)The time evolution of the $t\bar{t}$ system when the
effects of 
${\cal L}_{\mbox{\scriptsize {\it CP}-odd}}$ are included.
$\otimes$ represent the anomalous couplings.
(b)The potential $V_{\mbox{\scriptsize {\it CP}-odd}}$ tends to align
the $t$ spin in the direction of the color-electric
field and the $\bar{t}$ spin in the opposite direction.
      \label{cpodd}
}
  \hspace*{\fill}
\end{figure}

Now let us consider effects of the anomalous interactions 
eq.~(\ref{effop}) on the time 
evolution of the $t\bar{t}$ system (Fig.~\ref{cpodd}(a)).
{\it CP}-violation originating from the $t\gamma$ or $tZ$ coupling
occurs at the stage of the pair creation, 
i.e.\ when $t$ and $\bar{t}$ are
very close to each other.
The generated boundstate has $J^{PC}=1^{+-}$, so
$t$ and $\bar{t}$ are in the $P$-wave ($L=1$) and 
spin-0 state 
$\ket{\uparrow\downarrow}-\ket{\downarrow\uparrow}$.
On the other hand,
{\it CP}-violation 
originating from the $tg$ coupling takes place
after the boundstate formation when 
multiple gluons are exchanged between $t$ and $\bar{t}$,
i.e.\
when $t$ and $\bar{t}$
are separated at a distance of the Bohr radius.
The anomalous top-gluon coupling 
generates effectively a spin-dependent
potential between $t$ and $\bar{t}$
\bea
V_{\mbox{\scriptsize {\it CP}-odd}} = 
\frac{d_{tg}}{m_t} \, 
( \vc{s}_t - \bar{\vc{s}}_t ) \cdot \nabla V_{\rm C}(r) .
\label{cppot}
\eea
Here, 
$\vc{s}_t$ and $\bar{\vc{s}}_t$ 
denote the spins of non-relativistic
$t$ and $\bar{t}$, respectively; 
$V_{\rm C}(r) = - C_F {\alpha_s}/{r}$
is the Coulomb potential with the 
color factor $C_F = 4/3$.
When $d_{tg}>0$,
the potential $V_{\mbox{\scriptsize {\it CP}-odd}}$ tends to 
align both chromo-EDMs in the direction of chromo-electric field, or,
align $\bar{\vc{s}}_t$ in the direction of 
$\vc{r} = \vc{r}_t - \bar{\vc{r}}_{t}$
and $\vc{s}_t$ in the direction of $- \vc{r}$;
see Fig.~\ref{cpodd}(b).
Therefore, first the boundstate is formed in 
$J^{PC}=1^{--}$ ($L=0$ and $S=1$) state and after
interacting via the potential $V_{\mbox{\scriptsize {\it CP}-odd}}$
it turns into $J^{PC}=1^{+-}$
($L=1$ and $S=0$) state, i.e.\
the $t$ and $\bar{t}$ spins are aligned
into antiparallel directions.

We can disentangle the effects of the three couplings,
$d_{t\gamma}$, $d_{tZ}$, $d_{tg}$,
on the amplitude using the differences in the dependences on
the energy and $e^\pm$ polarization
in the threshold region.
Firstly when the c.m.\ energy is raised the {\it CP}-violating
effects due to $d_{t\gamma}$ and $d_{tZ}$ increase proportionally to
the velocity of the top quark, since these effects are induced directly by
the dimension-five operators.
On the other hand, the effect of $d_{tg}$ does not increase so
rapidly.
The enhancement of the $tg$-coupling
due to multiple exchanges of gluons will be lost
when the energy is raised and $t$ and $\bar{t}$ spread
apart quickly without forming boundstates.
Secondly, one may vary the relative weight of
the photon-induced {\it CP}-violating effect
and the $Z$-induced effect by varying 
the $e^\pm$ longitudinal polarization.
This is because $e_L$ and $e_R$
couple differently to $\gamma$ and $Z$, and
the relative weight of virtual $\gamma$ and $Z$ changes.

\subsection{{\it CP}-odd observables}

Which {\it CP}-odd observables are sensitive
to the {\it CP}-violating couplings $d_{t\gamma}$, $d_{tZ}$, $d_{tg}$?
For the process $e^+e^- \to t\bar{t}$, 
we may conceive of following 
expectation values of kinematical variables for
{\it CP}-odd observables:
\bea
&
\left< \,
( \vc{p}_{e} - \bar{\vc{p}}_{e} ) \cdot
( \vc{s}_t - \bar{\vc{s}}_t ) 
\, \right> ,
&
\nonumber \\
&
\left< \,
( \vc{p}_t - \bar{\vc{p}}_t ) \cdot
( \vc{s}_t - \bar{\vc{s}}_t )
\, \right> ,
&
 \\
&
\left< \,
[ ( \vc{p}_{e} - \bar{\vc{p}}_e ) \times
( \vc{p}_t - \bar{\vc{p}}_t ) ] \cdot
( \vc{s}_t - \bar{\vc{s}}_t )
\, \right> ,
&
\nonumber
\eea
where the spins and momenta are defined in the
c.m.\ frame.
(The initial state is {\it CP}-even
if we assume the SM interactions of $e^\pm$
with $\gamma$ and $Z$.)
The above quantities are the three components of the
difference of the $t$ and $\bar{t}$ spins.
One may easily confirm the {\it CP} transformations of the
above observables: e.g.\ 
$
(\vc{p}_t - \bar{\vc{p}}_t) \stackrel{C}{\rightarrow}
(\bar{\vc{p}}_t - \vc{p}_t) \stackrel{P}{\rightarrow}
(-\bar{\vc{p}}_t + \vc{p}_t)
$.
One might say (in a somewhat oversimplified way) that
in the SM the $t$ and $\bar{t}$ spins are parallel to each other, so
the SM contributions drop in the difference $ \vc{s}_t - \bar{\vc{s}}_t $,
whereas the $t$ and $\bar{t}$ spins become antiparallel to each other by
the effects of ${\cal L}_{\mbox{\scriptsize {\it CP}-odd}}$, so they
remain in $ \vc{s}_t - \bar{\vc{s}}_t $.
Thus, we want to
measure the difference of the spins of $t$ and $\bar{t}$.
It is equivalent to measuring the difference of the
polarization vectors of $t$ and $\bar{t}$.
All other {\it CP}-odd observables for $e^+e^- \to t\bar{t}$
are bilinear in $\vc{s}_t$ and $\bar{\vc{s}}_t$.
Since analyses of spin correlations are complicated, we focus on
the difference of the
polarization vectors.

Practically we can measure the $t$ and $\bar{t}$ polarization
vectors efficiently using $\ell^\pm$ angular distributions.
It is known that the
angular distribution of the charged lepton $\ell^+$ from the
decay of top quark is
maximally sensitive to the top quark polarization
vector.
In the rest frame of top quark,
the $\ell^+$ angular distribution is given by \cite{langdist}
\bea
\frac{1}{\Gamma_t} \,
\frac{d \Gamma ( t \to b \ell^+ \nu )}{d \cos \theta_{\ell^+}}
= \frac{1 + P \cos \theta_{\ell^+}}{2} 
\label{langdistr}
\eea
at tree level,
where $P$ is the top quark polarization and
$\theta_{\ell^+}$ is the angle of $\ell^+$ measured from the direction
of the top quark polarization vector.
Indeed the $\ell^+$ distribution is ideal for extracting
{\it CP}-violation in the $t\bar{t}$ {\it production process};
the above angular distribution is unchanged even if anomalous
interactions are included in the $tbW$ decay vertex,
up to the terms linear in the decay anomalous couplings and within the
approximation $m_b = 0$ \cite{new}.
Therefore, if we consider the average of the lepton direction,
for instance, we may extract the top quark
polarization vector efficiently:
\bea
\left< \vc{n} \cdot \vc{n}_\ell \right>_{\rm Lab}
\simeq \frac{1}{3} \, \vc{n}\cdot \vc{P} .
\eea
The average is to be taken at the top quark rest frame, but
in the threshold region, we may take the average in the
laboratory frame barely without loss of sensitivities to the
anomalous couplings.

\subsection{$t$ and $\bar{t}$ polarization vectors}

The polarization vectors of $t$ and $\bar{t}$ are defined from
the production cross section of a $t\bar{t}$ pair
in the threshold region.
The cross section, where ($t$,$\bar{t}$) have
momenta ($\vc{p}_t$,$-\vc{p}_t$) and the spins $+\frac{1}{2}$
along the quantization axes ($\vc{s}_t$,$\bar{\vc{s}}_t$)
in the c.m.\ frame, is given by
\bea
  \frac{d{\sigma}{}(\vc{s}_t,\bar{\vc{s}}_t)}{d^{3}{\vc{p}_t}}
  =   \frac{d{\sigma}{}}{d^{3}{\vc{p}_t}} \,
  \frac{%
      1
    + {\vc{P}}\cdot{\vc{s}_t}
    + {\bar{\vc{P}}}\cdot{\bar{\vc{s}}_t}
    + (\vc{s}_t)_i(\bar{\vc{s}}_t)_j\vc{Q}_{ij} }%
  {4}
  .
  \label{spinprojectedcs}
\eea
Here, $|\vc{s}_t|=|\bar{\vc{s}}_t|=1$.
On the right-hand-side,
$d\sigma/d^3{\vc{p}_t}$ represents 
the production cross section when the spins of
$t$ and $\bar{t}$ are summed over.
$\vc{P}$ and $\bar{\vc{P}}$ denote, respectively, the polarization
vectors of $t$ and $\bar{t}$.

According to the above definition we computed the 
polarization vectors.
The SM contributions to the polarization vectors are 
same for $t$ and $\bar{t}$, while the contributions from
${\cal L}_{\mbox{\scriptsize {\it CP}-odd}}$ are opposite in sign:
\bea
\vc{P}=\vc{P}_{\rm SM} + \delta \vc{P},
~~~~~~~~~
\bar{\vc{P}}=\vc{P}_{\rm SM} - \delta \vc{P} .
\eea
It is convenient to express $\delta \vc{P}$ in components:
\bea
\delta \vc{P} = \delta {\rm P}_\parallel \, \vc{n}_\parallel 
+ \delta {\rm P}_\perp \, \vc{n}_\perp
+ \delta {\rm P}_{\rm N} \, \vc{n}_{\rm N},
\eea
where the orthonormal basis is defined from the $e^-$ beam direction
and the top quark momentum direction as
\bea
\vc{n}_\parallel = \frac{\vc{p}_e}{|\vc{p}_e|},
~~~
\vc{n}_{\rm N} = \frac{\vc{p}_e \times \vc{p}_t}{|\vc{p}_e \times \vc{p}_t|},
~~~
\vc{n}_\perp = \vc{n}_{\rm N} \times \vc{n}_\parallel .
\eea
Then the {\it CP}-odd contributions are given by
\bea
&&
\delta {\rm P}_\parallel = 0,
\\
&&
\delta {\rm P}_\perp = 
{\rm Im} \Biggl[ \,
d_{tg} B^g_\perp \Bigl( \frac{D}{G} \Bigr) +
d_{t\gamma} B^\gamma_\perp \Bigl( \frac{F}{G} \Bigr) +
d_{tZ} B^Z_\perp \Bigl( \frac{F}{G} \Bigr) \,
\Biggr] 
\Bigl( \frac{p_t}{m_t} \Bigr) \sin \theta_t ,
\\
&&
\delta {\rm P}_{\rm N} = 
{\rm Re} \Biggl[ \,
d_{tg} B^g_{\rm N} \Bigl( \frac{D}{G} \Bigr) +
d_{t\gamma} B^\gamma_{\rm N} \Bigl( \frac{F}{G} \Bigr) +
d_{tZ} B^Z_{\rm N} \Bigl( \frac{F}{G} \Bigr) \,
\Biggr] 
\Bigl( \frac{p_t}{m_t} \Bigr) \sin \theta_t .
\eea
Here, $B^X_\perp$ and $B^X_{\rm N}$ denote combinations of the
electroweak couplings of $e^-$ and $t$ as well as of $e^\pm$
beam polarization.
$D$, $F$ and $G$ denote the QCD Green functions which incorporate
the boundstate effects.

\begin{figure}[tbp]
  \hspace*{\fill} \hspace{5cm}
    \includegraphics{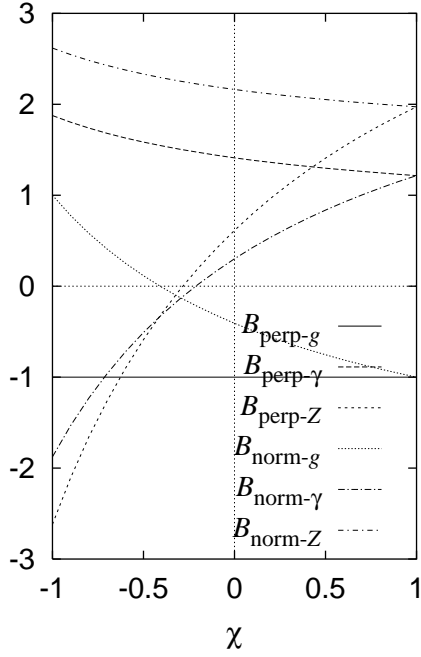}
  \hspace*{\fill}
  \caption{\small
      The electroweak coefficients $B_i^X$'s for $\delta {\rm P}_\perp$
      and $\delta {\rm P}_{\rm N}$
vs.\ the initial $e^\pm$ polarization parameter
      $\chi$.  
In the figure, 
      $B_{\rm para}=B_\parallel$, $B_{\rm perp}=B_\perp$, 
$B_{\rm norm}=B_{\rm N}$.
      \label{B}
  }
  \hspace*{\fill}
\end{figure}

In Fig.~\ref{B} we examine the electroweak coefficients 
$B^X_i$'s.
They are given as a function of the polarization parameter
of the initial $e^\pm$ beams
\bea
\chi = \frac{ P_{e^+} - P_{e^-} }{ 1 - P_{e^+} P_{e^-} } 
.
\label{chi}
\eea 
If the positron beam is unpolarized ($P_{e^+} = 0$), 
$\chi = - P_{e^-}$.
Typical sizes of all the coefficients $B^X_i$'s
are order one.
We also see that their dependences on the beam polarizations are
different.

\begin{figure}[tbp]
  \hspace*{\fill}
    \includegraphics[width=10cm]{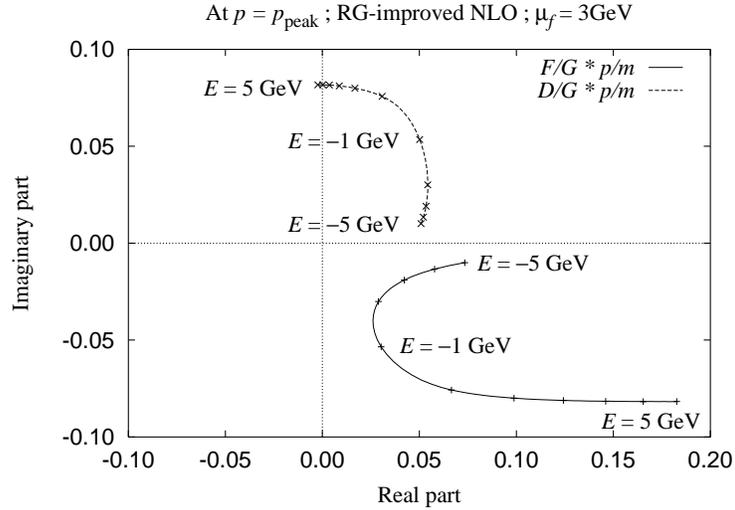}
  \hspace*{\fill}
  \caption{\small
      The ratios of the QCD Green functions times the velocity of 
      top quark evaluated at the 
      peak momentum $p_{\rm peak}$ of the momentum distribution.
      These are plotted on a
      complex plane as $E=\sqrt{s}-2m_t$ is varied.
      \label{complex_phase}
  }
  \hspace*{\fill}
\end{figure}

Next we examine the QCD factor
in Fig.~\ref{complex_phase}.
The ratios of the Green functions
together with the top quark velocity ($\beta = p_t/m_t$), 
$\beta D/G$ and $\beta F/G$, are shown on a
complex plane.
These are plotted
as a function of the energy 
$E=\sqrt{s}-2m_t$ alone by choosing
the top momentum to be the typical momentum at a
fixed energy.
As we raise the energy, the magnitude of the QCD factor associated
with the $t\gamma$ and $tZ$ couplings, $|\beta F/G|$, increase
proportionally to $\beta$.
On the other hand, the magnitude of the QCD factor associated
with the $tg$ coupling, $|\beta D/G|$, does not change very much.
We see that the size of $|\beta F/G|$ is 5--20\% while 
the size of $|\beta D/G|$ is 5--10\%.
Also it can be seen that the strong phases are quite sizable and
change rapidly with energy.
It is the characteristics of the $t\bar{t}$ boundstates that these
QCD factors can be computed reliably from the first principles.

\section{Conclusions}

In this work we studied how to probe the anomalous 
{\it CP}-violating couplings of the top quark with $\gamma$,
$Z$ and $g$ in the $t\bar{t}$ threshold region at
future $e^+e^-$ colliders.
The sensitivities to the anomalous couplings
are given and compared with other future experiments in Table~\ref{tab1}.
Qualitatively, the characteristics of the $t\bar{t}$ region are
summarized as follows.
\begin{enumerate}
\renewcommand{\labelenumi}{(\arabic{enumi})}
\setcounter{enumi}{0}
\item
We can measure the three couplings $d_{t\gamma}$, $d_{tZ}$, $d_{tg}$
simultaneously and we can disentangle each contribution.
\item
We can measure the complex phases of the couplings
$d_{t\gamma}$, $d_{tZ}$, $d_{tg}$.
Since the strong phases can be modulated at our disposal,
a single observable ($\delta {\rm P}_\perp$ or $\delta {\rm P}_{\rm N}$)
probes the phases of the couplings.
\item
Typical sizes of components of the 
difference of the $t$ and $\bar{t}$
polarizations are given by
$$
|\delta {\rm P}_\perp|, \, |\delta {\rm P}_{\rm N}| 
\sim ( \mbox{5--20}\% ) \times ( d_{t\gamma}, d_{tZ}, d_{tg} ) .
$$
They can be extracted efficiently from the directions of
charged leptons from decays of $t$ and $\bar{t}$:
$$
\langle \langle \vc{n} \cdot 
( \vc{n}_{\ell} + \bar{\vc{n}}_{\ell} )  \rangle \rangle _{\vc{p}_t}
\simeq
\frac{2}{3} \, 
\vc{n} \cdot \delta \vc{P} .
$$
\end{enumerate}

\begin{figure}[tbp]
  \hspace*{\fill}
    \includegraphics[width=10cm]{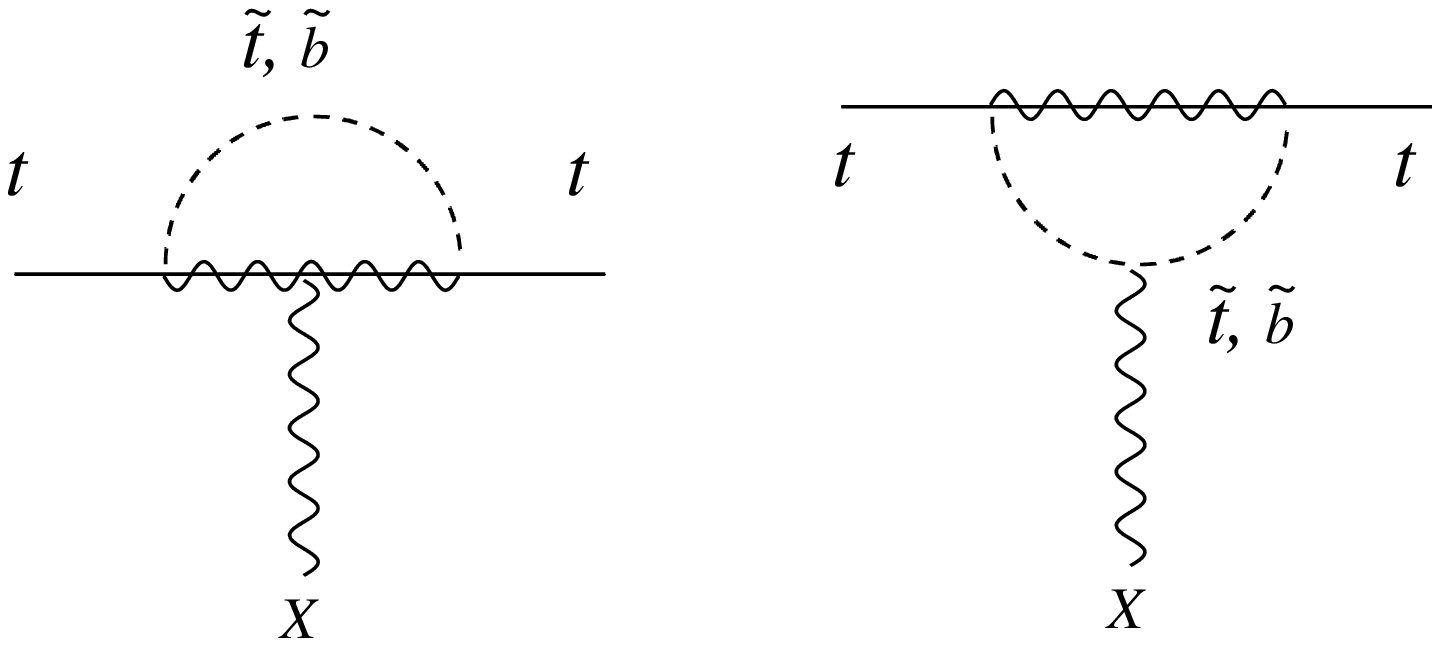}
  \hspace*{\fill}\\
\begin{center}
(a)\hspace{5.2cm}(b)
\end{center}
  \caption{\small
The diagrams which give rise to the top quark anomalous 
{\it CP}-violating couplings
in a supersymmetric model.
In diagram (a) only specific gaugino(s) couples to the gauge boson
$X = \gamma$, $Z$ or $g$.
On the other hand,
in diagram (b) all gauginos contribute.
      \label{susy}
  }
  \hspace*{\fill}
\end{figure}

We note that if one of the couplings is detected in the future, we 
would certainly want to measure the others 
in order to gain deeper understanding
of the {\it CP}-violating mechanism.
This is because one may readily think of 
various underlying processes which give different contributions
to the individual couplings;
see Fig.~\ref{susy}.

Regarding (1) and (2) above, QCD interaction is used as
a controllable tool for
the detection of the anomalous couplings.
This would be the first trial to use QCD interaction for such a
purpose without requiring any phenomenological inputs.

\section*{Acknowledgements}
This work is based on a collaboration with T.~Nagano and M.~Je\.zabek.
The author is grateful to the hospitality 
at the Summer Institute '99 (August 1999, Yamanashi, 
Japan) where this work
was initiated.
Also he thanks K.~Fujii, Z.~Hioki, K.~Ikematsu, T.~Takahashi, J.H.~K\"uhn,
S.~Rindani, M.~Tanabashi and M.~Yamaguchi 
for valuable discussions.
This work was supported in part
by the Japan-German Cooperative Science
Promotion Program.

\def\app#1#2#3{{\it Acta~Phys.~Polonica~}{\bf B #1} (#2) #3}
\def\apa#1#2#3{{\it Acta Physica Austriaca~}{\bf#1} (#2) #3}
\def\npb#1#2#3{{\it Nucl.~Phys.~}{\bf B #1} (#2) #3}
\def\plb#1#2#3{{\it Phys.~Lett.~}{\bf B #1} (#2) #3}
\def\prd#1#2#3{{\it Phys.~Rev.~}{\bf D #1} (#2) #3}
\def\pR#1#2#3{{\it Phys.~Rev.~}{\bf #1} (#2) #3}
\def\prl#1#2#3{{\it Phys.~Rev.~Lett.~}{\bf #1} (#2) #3}
\def\sovnp#1#2#3{{\it Sov.~J.~Nucl.~Phys.~}{\bf #1} (#2) #3}
\def\yadfiz#1#2#3{{\it Yad.~Fiz.~}{\bf #1} (#2) #3}
\def\jetp#1#2#3{{\it JETP~Lett.~}{\bf #1} (#2) #3}
\def\zpc#1#2#3{{\it Z.~Phys.~}{\bf C #1} (#2) #3}

\end{document}